\pdfoutput=1
\documentclass[a4paper,preprint,11pt,aps,prb,amsmath,amssymb,superscriptaddress]{revtex4-2}
\usepackage[T1]{fontenc}
\usepackage[utf8]{inputenc}
\usepackage[english]{babel}
\usepackage{lmodern}
\usepackage{amsfonts}
\usepackage{amsthm}
\usepackage{graphicx}
\usepackage{color}
\usepackage{xcolor}
\usepackage{url}
\usepackage{textcomp}
\usepackage{parskip}

\usepackage[defaultcolor=magenta,final]{changes}
\setauthormarkup{}
\definechangesauthor[name=MB, color=magenta]{MB}
\definechangesauthor[name=EB, color=blue]{EB}

\newcommand{\all}{Ti-Zr-Nb-(Cu,Ni,Co)}

\newcommand{\Ni}{$(\text{TiZrNbCu})_{1-x}\text{Ni}_{x}$}
\newcommand{\Cu}{$(\text{TiZrNbNi})_{1-x}\text{Cu}_{x}$}
\newcommand{\Co}{$(\text{TiZrNbCu})_{1-x}\text{Co}_{x}$}

\newcommand{\gmv}[1]{$\gamma = #1\,$mJ/mol$\,$K$^2$}
\newcommand{\gmvve}[2]{$(#1 \pm #2)\,$mJ/mol$\,$K$^2$}
\newcommand{\gmvv}[2]{$\gamma = (#1 \pm #2)\,$mJ/mol$\,$K$^2$}

\newcommand{\nefv}[1]{$N(E_F) = #1\,(\text{at eV})^{-1}$}

\newcommand{\nefvv}[2]{$N(E_F) = (#1 \pm #2)\,(\text{at eV})^{-1}$}
\newcommand{\susve}[1]{$#1\,$mJ/T$^2\,$mol}
\newcommand{\susv}[1]{$\chiexp = #1\,$mJ/T$^2\,$mol}
\newcommand{\susvve}[2]{$(#1 \pm #2)\,$mJ/T$^2\,$mol}

\newcommand{\pu}[1]{$#1$\%}
\newcommand{\pue}[2]{$#1$\%$\,$#2}

\newcommand{\Hc}{H_{c2}}
\newcommand{\Hcz}{\Hc(0)}
\newcommand{\xtot}{x_\text{tot}}
\newcommand{\chiexp}{\chi_\text{exp}}
\newcommand{\chiorb}{\chi_\text{orb}}
\newcommand{\lep}{\lambda_\text{ep}}

\renewcommand{\selectlanguage}[1]{}

\begin{document}
\title{Electronic structure property relationship in glassy \all\ alloys}
\thanks{Contact author: basletic@phy.hr}%

\author{Marko Kuveždić}
\affiliation{Department of Physics, Faculty of Science, University of Zagreb, Bijenička Cesta 32, 10000 Zagreb, Croatia}

\author{Emil Tafra}
\affiliation{Department of Physics, Faculty of Science, University of Zagreb, Bijenička Cesta 32, 10000 Zagreb, Croatia}

\author{Emil Babić}
\affiliation{Department of Physics, Faculty of Science, University of Zagreb, Bijenička Cesta 32, 10000 Zagreb, Croatia}

\author{Ramir Ristić}
\affiliation{Department of Physics, Faculty of Science, University of Zagreb, Bijenička Cesta 32, 10000 Zagreb, Croatia}

\author{Krešo Zadro}
\affiliation{Department of Physics, Faculty of Science, University of Zagreb, Bijenička Cesta 32, 10000 Zagreb, Croatia}

\author{Damir Starešinić}
\affiliation{Institute of Physics, Bijenička Cesta 46, HR-10000 Zagreb, Croatia}

\author{Ignacio A.~Figueroa}
\affiliation{Instituto de Investigaciones en Materiales, Universidad Nacional Autónoma de México (UNAM), Circuito Exterior s/n, Cd.~Universitaria, Ciudad de México 04510, {Mexico}}

\author{Mario Basletić}
\affiliation{Department of Physics, Faculty of Science, University of Zagreb, Bijenička Cesta 32, 10000 Zagreb, Croatia}

\date{\today}
\begin{abstract}
\replaced{In this work we revisit vast amount of existing data on physical properties of \all\ glassy alloys over a broad range of concentrations (from the high entropy range to that of conventional Cu-, Ni- or Co-rich alloys). 
By using our new approach based on total content of late transition metal(s), }
{By using a comprehensive study of physical properties of \all\ glassy alloys performed over a broad range of concentrations (from the high entropy range to that of conventional Cu-, Ni- or Co-rich alloys) }
we derive a number of physical parameters of a hypothetical amorphous TiZrNb alloy: lattice parameter $a = (3.42 \pm 0.02)$\AA, Sommerfeld coefficient \gmv{6.2}, density of states at \nefv{2.6}, magnetic susceptibility \susvve{2.00}{0.05}, superconducting transition temperature $T_c = (8 \pm 1)\,$K, upper critical field $\mu_0\Hcz = (20 \pm 5)\,$T, and coherence length $\xi(0) = (40 \pm 3)\,$\AA.
\replaced{We show that our extrapolated results for amorphous TiZrNb alloy would be similar to that of crystalline TiZrNb,}
{We show that the agreement between our extrapolated results for amorphous TiZrNb alloy and existing data for crystalline TiZrNb is quite good, } 
except for superconducting properties (most notably the upper critical field $\Hcz$), which might be attributed to the strong topological disorder of the amorphous phase.  
Also, we offer an explanation of the discrepancy between the variations of $T_c$ with the average number of valency electrons in \added[id=EB]{neighboring} alloys of 4d transition metals and some high entropy alloys. 
Overall, we find that our novel method of \replaced{systematic analysis of}{analyzing systematic} results is rather general, as it can provide reliable estimates of the properties of any alloy which has not been prepared as yet.
\end{abstract}

\maketitle

\section{Introduction}

In order to speed up the discovery of metallic alloys with novel advanced properties,alias a new alloy design based on multi-principal element solid solutions has been introduced at the beginning of this century \cite{miracle_high_2019}.
This alloy design was first applied to approximately equiatomic alloys of early (refractory)-late (TE-TL) transition metal alloys \cite{2002,cantor2002,KIM200317,ma16114031} in an attempt to discover new bulk metallic glasses (BMG) with the enhanced glass forming ability (GFA) \cite{INOUE2000279}.
A big advantage of this approach to fabrication of alloys, compared to the conventional one (based on one or two principal elements), is that it explores middle section of the multicomponent phase diagrams, thus allowing for an enormous number of new alloys and compounds available for research and potential applications.
Accordingly, this alloy design soon expanded to crystalline approximately equiatomic alloys with five or more principal components \cite{CANTOR2004213,Yeh2004} (so called high entropy alloys, HEA \cite{Yeh2004}, which are now a part of a broader familly of compositionally complex alloys, CCA \cite{ma16114031} which include the non-equiatomic alloys as well as those with fewer -- three or more -- principal components) and then to intermetallic and ceramic compounds, becoming a forefront of research in material science \cite{miracle_high_2019}.

Despite an early start \cite{2002,cantor2002,KIM200317}, the research of TE-TL CCAs (so called non-linear alloys in Ref.~\cite{miracle_high_2019}), both crystalline and amorphous (a-CCA) has progressed much less than that of CCAs composed mostly from the iron group of 3d-elements or refractory metals \cite{miracle_high_2019,senkov_accelerated_2015,Pickering2016,YE2016349,ZHANG20141,Guo2015,MIRACLE2017448,george_high-entropy_2019,IKEDA2019464,gao_high-entropy_2016,CANTOR2021100754}.
This is especially true for high-entropy metallic glasses (HEMG) which are still regarded as a new research field \cite{CHEN2021158852,LUAN202350}.
Despite modest research, the TE-TL HEAs showed some remarkable functional properties, including shape memory \cite{FIRSTOV2015S499,firstov_directions_2015,ma12244227,PIORUNEK2021157467,ma16083212} and thermoelectric performance \cite{adamo2024}, hydrogen storage capacity \cite{EDALATI2020387,ANDRADE202313555}, catalytic activity \cite{Qiu2019,glasscott_electrosynthesis_2019,glasscott_publisher_2019}, biomedical applications \cite{Popescu_2018,jfb13040263,met12111940}, electrical and superconducting performance \cite{ZHU2022118209,met10081078,Mizuguchi04032021}, and some other desirable properties \cite{han_multifunctional_2024}, but so far have found no practical applications, unlike these of a similar BMGs \cite{INOUE2024173546}.
In addition to properties relevant to potential applications, both TE-TL CCAs and conventional alloys show some features which are of considerable conceptual importance, as outlined below:\\
$i)$ Several TE-TL CCAs \cite{ma14195824,cunliffe_glass_2012,meng_phase_2019,NAGASE2018291,BabicAPL2024,park_phase_2016} and many conventional TE-TL alloys (e.g.~\cite{guntherodt_crystallization_1981,Ristic_2016,Babic14032018}) can be prepared in both the amorphous and crystalline states depending on the cooling rate of the molten alloys.
Thus, the study of alloys sharing this property could help to disentangle the role of topological and chemical disorder in forming their properties.
Moreover, the study of these alloys elucidated the variation of GFA with composition in conventional TE-TL alloys \cite{Ristic_2016,Babic14032018}.\\
$ii)$ All TE-TL alloys show the split-band structure of the electronic density of states (DOS) within the valence band (VB) \cite{ma14195824,BabicAPL2024,AMAMOU19801029,OELHAFEN19801017,ZEHRINGER1988317,solid-state_science_and_research_meeting_book_2017}.
The DOS close to the Fermi level $E_F$ is dominated by d-electrons of TEs whereas d-electrons of TLs are situated at higher binding energies, $E_B$ \cite{ma16041486}.
This results in a simple, often linear, variation of the DOS at $E_F$, $N(E_F)$, with TL content over considerable range of TL concentrations, which is reflected in a similarly simple variations of physical properties closely related to $N(E_F)$ (e.g.~\cite{ma14195824,ma16041486,RISTIC2015136}).
Furthermore, the position of $E_B$ and the width of the TL d-subbands depend on the TL content \cite{ma16041486}.
This results in the band crossing (the transition from the TE to TL dominated $N(E_F)$) which strongly affects the physical properties of TE-TL alloys \cite{ma14195824,ma16041486,KUVEZDIC2020119865,Ivkov_1984,PysRevB.33.3736,ma16041711}.
However, the split-band structure of the DOS makes the average number of valence electrons per atom (VEC) a poor approximation of the electronic structure (ES) of TE-TL alloys \cite{ma16041486,BABIC1981139}.\\
$iii)$ Many TE-TL alloys can be prepared by simple melt-spinning into a glassy state for TL concentrations \pu{20-70} \cite{RISTIC2015136,CHENG2011379}.
Such a broad glass forming range (GFR) facilitates the comparison between the models and experiment in binary alloys \cite{PhysRevB.27.2049,Jank_1991,HAFNER1992307,BAKONYI20052509,BAKONYI1995131}, and in corresponding multiprincipal element alloys it enables the study of the transition from a-HEAs to conventional metallic glasses with the same chemical make up \cite{ma14195824,ma16041486,KUVEZDIC2020119865,ma16041711,BILJAKOVIC20172661,babic_structure_2018,Ristic2019JAP,FIGUEROA2018455,Ristic2021JAP}.
Simultaneously, a broad GFR and a simple, often linear variations of properties with composition, enable a reliable extrapolation of the results for alloys to those of hypothetical pure amorphous TEs and TLs \cite{ma14195824,RISTIC2015136,BAKONYI20052509,BAKONYI1995131,babic_structure_2018,Ristic2019JAP,FIGUEROA2018455,Ristic2021JAP,Ristic2005FizA,RISTIC2007569,RISTIC2010S194,Tafra_2008}.
This method has been used in order to gain some insight into the properties of a pure amorphous Ti, Zr, Hf and Cu which have not been obtained in an amorphous state so far \cite{Han2024IntMat}.
We note that the method of accessing the properties of a pure amorphous element from the extrapolation of the results for its alloys is advantageous even in the case when this element has been obtained in the amorphous phase.
Namely, very high cooling rates $R_c\ge 10^{12}\,$K/s required to produce pure amorphous element results in the samples which are too small for measurement of any property other than the atomic structure \cite{Han2024IntMat}.
Further, such a high $R_c$ is likely to have a strong effect on the structure of an amorphous metal, which will differ from that obtained at lower $R_c$.

The same method\added{, studying class of physical properties dependence on conveniently chosen parameter,} can also be applied to obtain the information about alloys which have not been obtained in an amorphous phase so far.
Here we employ this method in order to obtain the physical properties of a \added{hypotetical amorphous} TiZrNb refractory alloy from the novel type of analysis of the results for glassy \all\ alloys \cite{ma14195824,KUVEZDIC2020119865,ma16041711,babic_structure_2018,Ristic2019JAP,FIGUEROA2018455,Ristic2021JAP}.
\deleted{( }Our analysis involves the extrapolations of the variations of physical properties with the total content of TLs, \added{denoted by $\xtot$,} and the use of nonlinear representations of these variations when appropriate. \deleted{)}
We compare the properties of amorphous TiZrNb alloy with those of the same alloy in the crystalline phase with body centered cubic structure \cite{met11111819}.
The TiZrNb alloy is very important since it forms the basis for the majority of refractory HEAs and superalloys \cite{SENKOV2018201}.
This alloy is also a very promising biomedical material \cite{HU2022107725}.
In biomedical applications the amorphous phase of this alloy would outperform the crystalline one due to its higher strength and hardness, and lower corrosion and wear.
\added[id=EB]{Indeed, we will show that while most of the derived properties of glassy TiZrNb alloy seems to be similar to its crystalline counterpart (which is probably due to similar local atomic arrangements and thus ES in two phases), some of its transport and superconducting properties, important for application, seem probably better \cite{met11111819,Kosut2024LTP}.}


We note that the application of the method employed in our study of the properties of the hypothetical TiZrNb glassy alloy is not limited to amorphous systems only, and can be applied equally well to crystalline alloy systems.
As an example we note that the properties of Fe--rich (CrMnCoNi)$_{1-x}$Fe$_x$ alloys with the single face centered cubic structure, fcc, which cannot be synthesized with standard metallurgical methods for $x>0.5$ \cite{BRACQ2017327}, can be determined from the extrapolation of results for $x\le 0.5$ to $x> 0.5$.

\section{Experimental Details}

The materials, methods of preparation and characterization of the samples and the techniques of the measurements used for \all\ glassy alloys were reported earlier \cite{ma14195824,ma16041486,KUVEZDIC2020119865,ma16041711,babic_structure_2018,Ristic2019JAP,Ristic2021JAP}.
For completeness, we briefly review some of these issues.

The ingots weighing about gram of seven \Co\ alloys ($x = 0$, 0.19, 0.20, 0.25, 0.32, 0.43 and 0.50 \cite{Ristic2021JAP}), ten \Cu\ alloys ($x=0$, 0.05, 0.12, 0.15, 0.20, 0.25, 0.32, 0.35, 0.43 and 0.50 \cite{Ristic2019JAP}) and eight \Ni\ alloys ($x = 0$, 0.125, 0.15, 0.2, 0.25, 0.35, 0.43 and 0.50 \cite{babic_structure_2018}) were prepared from high-purity metals ($\ge 99.8\,$at.$\,$\%) by arc melting in high-purity argon and in the presence of a titanium getter.
From ingots thin ribbons with a thickness of about $25\,\mu$m and a width of $2\,$mm were prepared by melt spinning molten alloys on the surface of the cooper roller rotating at $25\,$m/s in a pure helium atmosphere.
The X-ray diffraction (XRD) patterns with Cu-K$\alpha$ radiation showed that all samples, except those with $x=0\,$Co and $x = 0.50\,$Co, were X-ray amorphous \cite{babic_structure_2018,Ristic2019JAP,Ristic2021JAP}.
The amorphous samples were further characterized with differential scanning calorimetry and thermogravimetric analysis which provided the thermal parameters and confirmed the amorphous state of as-cast samples \cite{ma14195824,babic_structure_2018,Ristic2019JAP,Ristic2021JAP}.
The chemical characterization (scanning electron microscopy with energy dispersive spectroscopy (SEM/EDS)) confirmed a random distribution of constituent elements down to a micrometer scale, and the calculated average concentrations were within $1\,$at.$\,$\% of nominal \cite{babic_structure_2018,Ristic2019JAP,FIGUEROA2018455}.

We briefly revisit the measurements of the properties which will be used for the determination of the corresponding properties of amorphous TiZrNb alloy.
The magnetic susceptibility of all alloys was measured with a Quantum Design magnetometer, MPMS 5, in a magnetic field $B$ up to $5\,$T and the temperature range $2-300\,$K \cite{babic_structure_2018,Ristic2019JAP,Ristic2021JAP}.
Since the magnetic susceptibility of all alloys, except for that with $x=0.5\,$Co, showed a weak dependence on temperature (as is usual for nonmagnetic TE-TL alloys, e.g.\ \cite{RISTIC2015136,BABIC1981139,babic_structure_2018,Ristic2019JAP,Ristic2021JAP}) we use the room temperature values in the following analysis.
The measurements of the low temperature specific heat, LTSH on \Ni\ and \Cu\ alloys, and TiZrNbCuCo sample were performed in the temperature range $1.8-300\,$K using a Physical Property Measurement System, PPMS Model 5000 from Quantum Design, as described in more details elsewhere \cite{BILJAKOVIC20172661,babic_structure_2018,Ristic2019JAP,Ristic2021JAP}.

The low temperature resistive superconducting transition measurements down to $300\,$mK were performed in a He$^3$ insert (Heliox, Oxford Instruments) of a $16/18\,$T Oxford superconducting magnet system, in the magnetic fields up to $\pm 4\,$T.
The electrical resistivity was measured using a low frequency ($22\,$Hz) AC method with rms currents in the $20-200\,\mu$A range, while voltages were measured with a dual-phase Signal Recovery 5120 Lock-in amplifier.
The magnetic field was oriented perpendicular to the broad surface of the ribbon samples and to the current direction. The temperature was measured with a Lake Shore 340 Temperature Controller with a calibrated Cernox thermometer situated close to the sample \cite{KUVEZDIC2020119865,ma16041711,Tafra_2008}.

\section{Results and Discussion}

\subsection{Lattice parameters $a$}

The atomic structure of \all\ glassy alloys has been investigated in some detail, both by XRD patterns with a Cu-K$\alpha$ source and the synchrotron based high-energy X-ray diffraction (HEXRD) at the Diamond Light Source with a monochromatic beam of $0.01545\,$nm wavelength \cite{ma14195824,babic_structure_2018,Ristic2019JAP,Ristic2021JAP}.
There is a quite good, quantitative, agreement between the average near neighbor atomic distances, $d$, determined from the XRD patterns by using the approximate Guinier expression \cite{Calvayrac01101983}
\begin{equation}
    \label{eq:Guinier}
    d = \frac{7.73}{k_p},
\end{equation}
where $k_p$ is the modulus of scattering vector corresponding to the first maximum in the diffraction pattern, and those from the positions of the first maxima of the radial distribution functions $R(r)$ obtained form the HEXRD measurements \cite{ma14195824,babic_structure_2018,Ristic2019JAP,Ristic2021JAP}.

A large difference between the atomic radii of the employed TE and TL elements of \all\ alloys and their strong interatomic interactions are expected to result in a bcc like local atomic arrangements \cite{ma14195824,BILJAKOVIC20172661,Ristic2019JAP,Ristic2021JAP} resembling those in similar crystalline TE-TL HEAs and CCAs \cite{FIRSTOV2015S499,ma16083212,ma14195824,cunliffe_glass_2012,meng_phase_2019,NAGASE2018291,BabicAPL2024,park_phase_2016}.
(This conjecture is supported with the appearance of a dominant bcc phase in bulk crystalline \Ni\ samples with $x\le 0.2$ \cite{cunliffe_glass_2012} and the observation of a bcc phase upon crystallization of our glassy sample with $x=0.125$ \cite{ma14195824}.) 
Thus, assuming a bcc-like local atomic structure we estimated the corresponding lattice parameters $a = 2d/3^{0.5}$ for all our alloys \cite{ma14195824,BILJAKOVIC20172661,babic_structure_2018,Ristic2019JAP,FIGUEROA2018455,Ristic2021JAP}.
These estimated values of $a$ agreed quite well with those calculated from the lattice parameters for the bcc phases of constituents by assuming the validity of the Vegard's law, except for the alloy with $x=0.5\,$Ni and some Co containing alloys \cite{ma14195824,babic_structure_2018,Ristic2021JAP}.
(The estimated values were however a little larger than corresponding values calculated using Vegard's law probably due to somewhat less dense atomic packaging in the amorphous state \cite{ma14195824,BILJAKOVIC20172661,babic_structure_2018,Ristic2019JAP,FIGUEROA2018455,Ristic2021JAP}.) 
The observed deviations of $a$ from a linear Vegard's law-like variation with TL content in $0.5\,$Ni and some Co containing alloys were accompanied with an increase in the average number of the nearest neighbor atoms (calculated from $R(r)$) and are associated with the band crossing \cite{ma16041486} as confirmed by the rapid change of the Hall coefficient in these alloys \cite{KUVEZDIC2020119865,ma16041711}.
To our knowledge this was the first observation of the effect of band crossing on the interatomic distance and the local atomic configuration.

\begin{figure}
\centering
\includegraphics[width=0.7\linewidth]{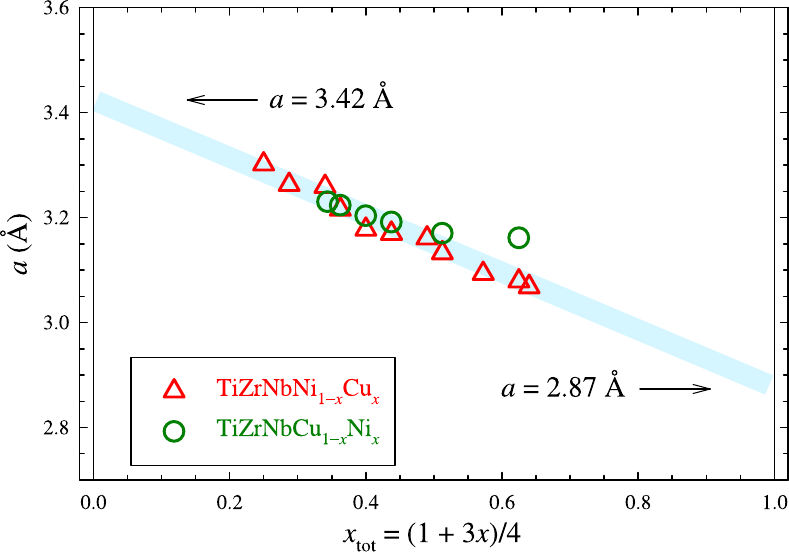}
\caption{ 
The lattice parameter $a$ of \Cu\ and \Ni\ glassy alloys (red triangles and green circles, respectively) for various $x$.
Thick light blue line denotes linear extrapolation to $\xtot \rightarrow 0$, and $\xtot \rightarrow 1$, i.e.\ to glassy TiZrNb alloy and amorphous Cu, respectively.
Data are adapted from Refs.\ \cite{ma14195824,Ristic2019JAP,FIGUEROA2018455,Ristic2021JAP}.
Arrows denote extrapolated values of lattice parameter $a = 3.42\,$\AA\ and $a = 2.87\,$\AA, for $\xtot = 0$ and $\xtot = 1$, respectively.\label{fig:abcc-vs-x_tot}
}
\end{figure}

As seen in Figure \ref{fig:abcc-vs-x_tot} the values of $a$ of all \Cu\ and \Ni\ glassy alloys (except for that with $x=0.5\,$Ni) decrease linearly with total content of Cu and Ni, $\xtot = (1 + 3x)/4$\added{, where $\xtot$ is total concentration of late transition metal(s) Cu and Ni}.
This variation extrapolates to $a = (3.42 \pm 0.02)$\AA\ for $\xtot = 0$, i.e.\ the amorphous TiZrNb alloy.
The estimated value of $a$ of amorphous TiZrNb alloy is about \pu{0.5} larger than that of the bcc phase of the corresponding crystalline alloy ($a = 3.404\,$\AA\ in Ref.\ \cite{met11111819}).
As explained above somewhat larger $d$ (hence also $a$) in amorphous alloy is expected due to somewhat less dense atomic packing in the amorphous phase.

The representation of the composition dependence of the structure of an alloy in terms of the average atomic volume, $V_a$, has some advantage compared to that using $a$ \cite{RISTIC2015136,BAKONYI20052509,babic_structure_2018,Ristic2019JAP,FIGUEROA2018455,Ristic2021JAP}.
In the bcc structure $V_a=a^3/2$ which yields $V_a =(20.00\pm 0.35)\,$\AA$^3$ for the average atomic volume in the amorphous TiZrNb alloy.
This value is as expected a little larger than that of the crystalline TiZrNb, $V_a = 19.72\,$\AA$^3$.
One advantage of using $V_a$ instead of $a$, is that $V_a$ can be independently determined from the measured mass density $D$ without any prior assumption about the atomic structure of the alloys \cite{RISTIC2015136,BAKONYI20052509}:
\begin{equation}
    \label{eq:V_a}
    V_a = 1.66\, M_a/D,
\end{equation}
where $M_a$ is the composition-averaged molar mass of the alloy.
(We note that the knowledge of $V_a$ is necessary to determine the average local atomic packing fraction in an alloy \cite{RISTIC2015136,BAKONYI20052509,babic_structure_2018,Ristic2019JAP,Ristic2021JAP}.)
The expression (\ref{eq:V_a}) can also be used in order to determine the $D$ of an alloy from the known $V_a$ and $M_a$.
From $V_a = 20.00\,$\AA$^3$ and $M_a = 77.33\,$g/mol we obtain $D=(6.42\pm 0.12)\,$g/cm$^3$ for amorphous TiZrNb alloy, which is considerably lower than that reported for crystalline TiZrNb ($6.63\,$g/cm$^3$, \cite{met11111819}).
We note however that the TiZrNb alloy studied in Ref.\ \cite{met11111819} was somewhat non-equiatomic with some excess of Nb and deficiency of Ti.
There is also a discrepancy between the value of $D$ reported in \cite{met11111819}, and that calculated using Eq.\ (\ref{eq:V_a}), $D = 6.51\,$g/cm$^3$, which may indicate some error in the measurement of $D$ in \cite{met11111819}.
Since the variation of $a$ in Figure \ref{fig:abcc-vs-x_tot} is dominated with the results for \Cu\ alloys which showed the ideal solution behaviour extrapolating to the value for pure Cu for $x=1$, we expect the same result from extrapolation of the variation of $a$ in Figure \ref{fig:abcc-vs-x_tot} to $\xtot = 1$.
From the extrapolated $a = (2.87\pm 0.02)\,$\AA\ we obtain $V_a = (11.82\pm 0.2)\,$\AA$^3$ which is practically the same as that determined from the ideal solution behaviours in binary \cite{RISTIC2015136} and quinary \cite{Ristic2019JAP} TE-Cu amorphous alloys.
A meaningful values of $a$, $V_a$ and $D$ obtained by extrapolations of the variations of $a$ with $\xtot$ in Figure \ref{fig:abcc-vs-x_tot} to both $\xtot = 0$ and $\xtot = 1$ seem to lend some support to this approach in analyzing the physical properties of our alloys.

\subsection{Somerfeld coefficient $\gamma$ and density of states $N(E_F)$}
\begin{figure}
\includegraphics[width=0.5\linewidth]{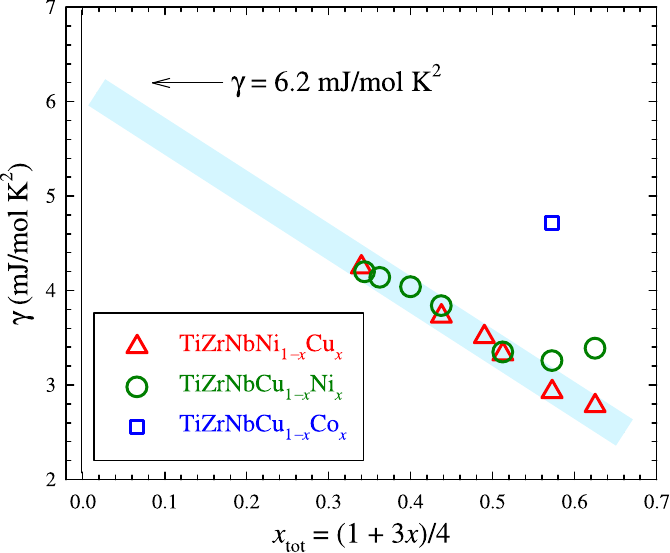}
\includegraphics[width=0.5\linewidth]{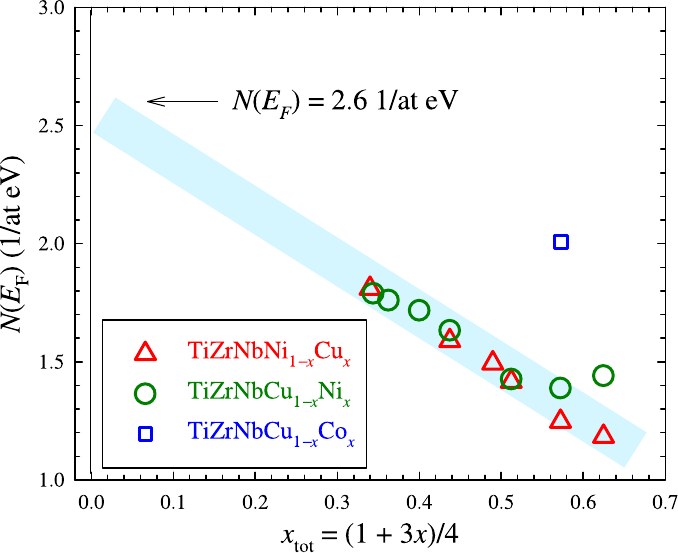}
\caption{
The Sommerfeld coefficient $\gamma$ (left panel) and dressed density of states $N(E_F)$ at $E_F$ (right panel) of \Cu, \Ni\ and \Co\ glassy alloys (red triangles, green circles and blue squares, respectively) for various $x$.
Data are adapted from Refs.\ \cite{ma14195824,babic_structure_2018,BILJAKOVIC20172661,FIGUEROA2018455,Ristic2021JAP}.
Left and right panel: thick light blue line denotes linear extrapolation to $\xtot \rightarrow 0$, i.e.\ extrapolation to glassy TiZrNb alloy.
Arrows denote extrapolated values for \gmv{6.2}, and \nefv{2.6}.\label{fig:gamma and nef}
}
\end{figure}

Since the ES determines all intrinsic properties of alloys and materials \cite{Kittel2005,QI201911} its knowledge is especially important for the selection of alloys with desirable properties \cite{IKEDA2019464,ma14195824}.
Moreover, most of the physical properties of the alloys depends on $N(E_F)$ \cite{ma14195824,KUVEZDIC2020119865,ma16041711,Kittel2005,Dynes_1976}, thus it is often sufficient to measure LTSH in order to explain these properties.
Unfortunately, there is still lack of systematic LTSH measurements in HEAs and CCAs, which is detrimental both for their understanding and practical applications \cite{ma14195824,ma16041486,BILJAKOVIC20172661,babic_structure_2018,Ristic2021JAP}.
We note however that correct determination of $N(E_F)$ of glassy alloys from LTSH measurements is non-trivial \cite{Remenyi2014APL} since in addition to the Debye term in the vibrational contributions to LTSH one has also to account for the contribution associated with the so-called boson peak \cite{babic_structure_2018,Ristic2019JAP,FIGUEROA2018455,Ristic2021JAP}.
The electronic contribution to LTSH is contained in the Sommerfeld coefficient $\gamma$ of a term linear in temperature \cite{Kittel2005,Remenyi2014APL}:
\begin{equation}
    \label{eq:gamma}
    \gamma = \pi^2k_B^2N(E_F)/3
\end{equation}
where $k_B$ is the Boltzmann constant and $N(E_F)$ is the (dressed) DOS at $E_F$ which is enhanced with respect to the bare (band) DOS at $E_F$, $N(E_F) = (1 + \lep)N_0(E_F)$ with $\lep$ the electron-phonon coupling constant.
Thus, variation of $\gamma$ directly reflects that of $N(E_F)$ as is seen in Figure \ref{fig:gamma and nef} which shows both $\gamma$ and $N(E_F)$ for all \all\ glassy alloys for which we measured LTSH \cite{ma14195824,babic_structure_2018,Ristic2019JAP,FIGUEROA2018455,Ristic2021JAP}.
Both $\gamma$ and $N(E_F)$ for all alloys (except for those with \pue{43}{Co} and \pue{50}{Ni} which are above the crossover concentration for the band crossing \cite{ma16041486}) follow a linear variations with $\xtot$ which extrapolate to \nefv{2.6}\ and \gmv{6.2}\ for the amorphous TiZrNb alloy.
To our knowledge there are no results for $\gamma$ of the crystalline TiZrNb alloy \cite{met11111819,Kosut2024LTP}, but rather high values of $N_0(E_F)$ were predicted for the bcc phase of TEs \cite{BakonyiPRB1993}.
These extrapolated values are consistent with high values of \gmvv{6.2}{0.2}\ and \gmvve{7.9}{0.3}\ for amorphous Zr and Ti, respectively, obtained from the extrapolation of results for amorphous Zr(Ti)-Cu alloys to zero Cu content \cite{RISTIC2015136}.
The value of $\gamma$ of a pure (bcc) Nb is also high, \gmvve{7.60}{0.02}\ \cite{SonierPRB2006}, partially due to large $\lep$ of Nb.
High value of $N(E_F)$ of amorphous TiZrNb alloy is expected to result in a rather high superconducting transition temperature $T_c$ \cite{RISTIC2010S194,Tafra_2008,Kosut2024LTP}, an enhanced magnetic susceptibility \cite{RISTIC2015136,Ristic2005FizA,RISTIC2007569,RISTIC2010S194,Tafra_2008}, and not that high elastic moduli, strength and hardness \cite{QI201911,Ristic11122007}.
Since the ideal solution behaviour affects all properties of an alloy system we expect that the results for $\gamma$ and $N(E_F)$ of our alloys (Figure \ref{fig:gamma and nef}) extrapolate to those of a pure Cu for $\xtot = 1$.
Indeed, the extrapolation of the data in Figure \ref{fig:gamma and nef} to $\xtot = 1$ yields \gmvv{0.65}{0.2}\ and \nefvv{0.3}{0.2}\ which are -- within the error -- the same as those obtained from the study of Ti(Zr)-Cu glassy alloys \cite{RISTIC2015136}, and they agree quite well with these of a pure crystalline Cu.

\subsection{Magnetic susceptibility $\chiexp$}
\begin{figure}
\centering
\includegraphics[width=0.6\linewidth]{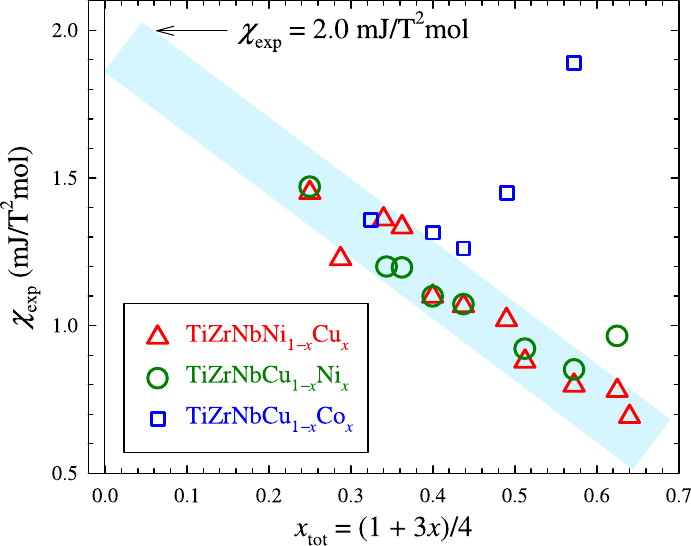}
\caption{ 
The magnetic susceptibility $\chiexp$ of \Cu, \Ni\ and \Co\ glassy alloys (red triangles, green circles and blue squares, respectively) for various $x$.
Data are adapted from Refs.\ \cite{ma14195824,babic_structure_2018,Ristic2019JAP,Ristic2021JAP,BILJAKOVIC20172661,FIGUEROA2018455}.
Thick light blue line denotes linear extrapolation to $\xtot \rightarrow 0$, i.e.\ to glassy TiZrNb alloy.
Arrow denote extrapolated values of magnetic susceptibility $\chiexp = 2.0\,$mJ/T$^2\,$mol.\label{fig:suscep}
}
\end{figure}

A close relationship between the ES and the magnetic properties of TE-TL alloys is shown clearly by the variations of the room temperature magnetic susceptibility, $\chiexp$, of all \all\ glassy alloys with composition.
The data in Figure \ref{fig:suscep} show that similar to $\gamma$ and $N(E_F)$ (see Figure \ref{fig:gamma and nef}), most values of $\chiexp$ decrease approximately linearly with increasing $\xtot$. \added{(Here the $\xtot$ denotes total concentration of late transition metal(s) Cu, Ni and Co.)}
The exceptions are the values of $\chiexp$ for the alloys with \pue{50}{Ni}, and \pu{32} and \pue{43}{Co} which are influenced by the effects of the band crossing \cite{ma16041486,babic_structure_2018,Ristic2021JAP}.
The variation of $\chiexp$ in Figure \ref{fig:suscep} extrapolates to \susvve{2.00}{0.05}\ and \susvve{-0.07}{0.05}\ for $\xtot = 0$ and $1$, respectively.
The latter value of $\chiexp$ ($\xtot = 1$) is within the extrapolation error the same as that of a pure (crystalline) Cu \cite{RISTIC2015136}.
To our knowledge there is no measurement of $\chiexp$ for TiZrNb alloy ($\xtot = 0$), but our result compares favourably with the corresponding values of its components: \susv{2.14}\ for pure Nb \cite{COLLINGS1972389}, and \susv{2.33}\ and \susve{1.56}\ which were obtained from extrapolation of the results for glassy Ti,Zr-Cu alloys to pure Ti and Zr, respectively \cite{RISTIC2015136}.
Moreover, the average of these values, $\chi_\text{av} = 2.01\,$mJ/T$^2\,$mol is almost the same as our extrapolated value of $\chiexp$ for $\xtot = 0$.

Despite very similar variation with $\xtot$ as that of $\gamma$ and $N(E_F)$, $\chiexp$ of transition metals and their alloys is quite complex \cite{RISTIC2015136,BakonyiPRB1993}:
\begin{equation}
    \label{eq:chiexp}
    \chiexp = \chi_P + \chiorb + \chi_\text{dia},
\end{equation}
where $\chiorb$ and $\chi_\text{dia}$ are, respectively, the orbital paramagnetic and diamagnetic contributions obtained by adding the corresponding contributions from constituents.
The Pauli paramagnetism of the d-band $\chi_P$ is enhanced over the free electron value $\chi_P^0 = \mu_B^2N_0(E_F)$, where $\mu_B$ is the Bohr magneton, by the exchange interaction: $\chi_P = S\chi_P^0$, where $S>1$ is the Stoner enhancement factor \cite{BakonyiPRB1993}.
Thus, the direct contribution of $N_0(E_F)$ to $\chiexp$ in TE-TL alloys is quite moderate and the main contribution in TE rich alloys is $\chiorb$ \cite{RISTIC2015136}.
Indeed, the composition-averaged $\chiorb$ in TiZrNb alloy is \susve{1.12}\ \cite{PlacePSS1971}, thus accounts for \pu{56} of $\chiexp$, and the decrease of $\chiorb$ with increasing $\xtot$ dominates the variations of $\chiexp$ of majority of alloys in Figure \ref{fig:suscep}.

\subsection{Superconducting transition temperature $T_c$}
The research of HEA superconductors (mostly refractory metal based alloys) has expanded rapidly \cite{met10081078} since the discovery of the Ta-Nb-Hf-Zr-Ti superconductor with the superconducting transition temperature $T_c = 7.3\,$K in 2014 \cite{PhysRevLett.113.107001}.
Despite $T_c$s similar to or lower than those of conventional practical superconductors \cite{met10081078,Mizuguchi04032021, Kosut2024LTP,PhysRevLett.113.107001,SunPRM2019} they often show some properties such as the strong coupling behaviour \cite{KIM2020250}, high upper critical field $\Hc$ \cite{Kosut2024LTP,Li_2024}, high critical current density $J_c$, irradiation resistance \cite{jung_high_2022} and stability up to very high pressures \cite{SunPRM2019}, which are important for the practical applications of superconductivity.

The appearance of superconductivity and the value of $T_c$ are closely related to the ES of an alloy.
Moreover, in disordered transition metals and their alloys \cite{KUVEZDIC2020119865,Tafra_2008} the superconductivity seems to be governed by $N_0(E_F)$, which results in an approximately linear variation of $\lep$ with $N_0(E_F)$, with a slope that depends only on the class of transition metals (3d, 4d or 5d) and the position of $E_F$ within a given d-band DOS \cite{Dynes_1976}.
In TE-TL alloys this leads to a decrease of the value of $T_c$ on increasing TL content (thus also increasing the value of average VEC) irrespective of whether these alloys are crystalline or amorphous, binary or HEA \cite{KUVEZDIC2020119865,ma16041711,BABIC1981139,RISTIC2010S194,Tafra_2008,SunPRM2019,XU2024115986,AltounianPRB1983,KarkutPRB1983,StolzeCoM2018,babic_conference_1982}.

\begin{figure}
\centering
\includegraphics[width=0.6\linewidth]{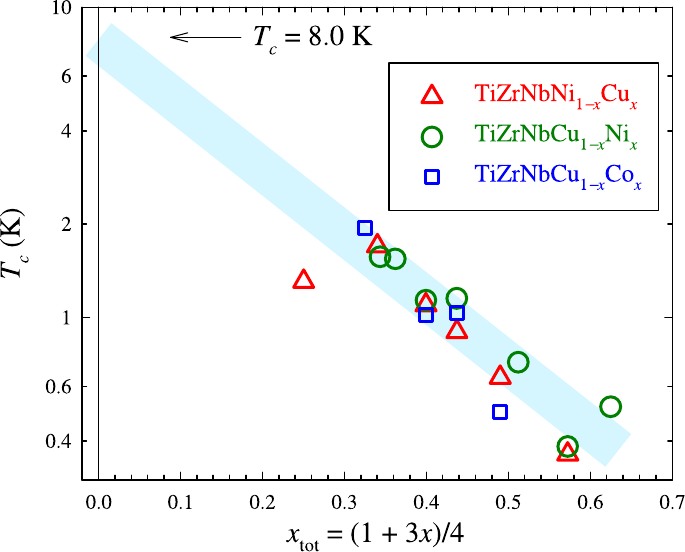}
\caption{ 
The superconductivity transition temperature $T_c$ of \Cu, \Ni\ and \Co\ glassy alloys (red triangles, green circles and blue squares, respectively) for various $x$.
Data are adapted from Refs.\ \cite{KUVEZDIC2020119865,ma16041711}.
Thick light blue line denotes linear extrapolation to $\xtot \rightarrow 0$, i.e.\ to glassy TiZrNb alloy.
Arrow denote extrapolated value of transition temperature $T_c = 8.0\,$K.\label{fig:Tc}
}
\end{figure}

In Figure \ref{fig:Tc} we show the variation of $T_c$s of glassy \all\ alloys \cite{KUVEZDIC2020119865,ma16041711} with $\xtot$.
Note a logarithmic scale for $T_c$ which is suitable for an exponential variation.
Such a plot of $T_c$ vs.\ $\xtot$ is plausible since in these alloys $N(E_F)$ varies linearly with $\xtot$ (see Figure \ref{fig:gamma and nef}) and the BCS theory of superconductivity \cite{tinkham2004introduction,McMillanPR1968} predicts in both, the weak and strong coupling regime, approximately exponential variation of $T_c$ with $N(E_F)$.
Indeed, within the experimental scatter, the data for $T_c$ of the majority of our alloys follow an approximately exponential variation.
The exception are the values of $T_c$ for alloys with \pu{43} of Co and \pu{50} of Ni, as well as those for \pu{0} and \pu{50} of Cu, which fall below the (error) band determined by the most data in Figure \ref{fig:Tc}.
As already explained for $\chiexp$ (Figure \ref{fig:suscep}) a little deviation of $T_c$ data for the alloys with \pue{32}{Co} and \pue{50}{Ni} from the common trend of the data in Figure \ref{fig:Tc} is due to band crossing, thus an enhancement of the contribution of d-states of Co and Ni to $N(E_F)$ at these compositions \cite{ma16041486}.
A large deviation of $T_c$ of the alloy with \pue{0}{Cu} ($\xtot = 0.25$) from that predicted by the common trend of the data is due to extraordinary sensitivity of $T_c$s of TE-TL glassy alloys on incipient crystallization \cite{BABIC1981139,Tafra_2008,AltounianPRB1983}.
A small suppression of $T_c$ of the alloy with \pue{50}{Cu} ($\xtot = 0.625$) is probably due to a small local chemical inhomogeneity (such as small increase in Cu content in a small part of that particular sample).
As already noted \cite{KUVEZDIC2020119865} the values of $T_c$ of alloys in Figure \ref{fig:Tc} are quite low ($T_c \le 1.9\,$K) compared to these of Zr-Cu,Ni glassy alloys \cite{BABIC1981139,Tafra_2008,AltounianPRB1983,KarkutPRB1983}.
Somewhat lower $N(E_F)$ values of our alloys \cite{ma14195824,KUVEZDIC2020119865}, together with the adverse influence of Ti \cite{ma16041711,Tafra_2008,Dynes_1976} probably explain their low $T_c$s.
Recent study of the influence of the increase in Ti, Zr or Nb content of $T_c$s of Ti-Zr-Nb-Cu-Ni alloys has shown that Ti$_{0.3}$Zr$_{0.15}$Nb$_{0.15}$Cu$_{0.2}$Ni$_{0.2}$ alloy despite of higher $N(E_F)$ has nearly \pu{40} lower $T_c$ than these of Zr$_{0.3}$Ti$_{0.15}$Nb$_{0.15}$Cu$_{0.2}$Ni$_{0.2}$ and Nb$_{0.3}$Ti$_{0.15}$Zr$_{0.15}$Cu$_{0.2}$Ni$_{0.2}$ alloys \cite{ma16041711}.

The variation of $T_c$s in Figure \ref{fig:Tc} extrapolates for $\xtot = 0$ to $T_c = (8 \pm 1)\,$K for glassy TiZrNb alloy which agrees very well with that of crystalline (bcc) TiZrNb alloy, $T_c = 8.2\,$K \cite{Kosut2024LTP}.
For better comparison with $T_c$ of crystalline alloy we extrapolated $T_c$s defined as the midpoints of the resistive transition in zero applied field.
If we used some more rational definition of $T_c$s (free from the effects of flux flow and superconducting fluctuations \cite{KUVEZDIC2020119865}) our extrapolated value of $T_c$ would be a little higher.
The value of superconducting gap determined from the point-contact Andreev reflection measurements indicates a weak coupling superconductivity of bcc TiZrNb alloy \cite{Kosut2024LTP}
.
Unfortunately, a scatter in the values of the Debye temperature ($\theta_D$) of our alloys prevents a reliable extrapolation of these values to $\xtot =0$, i.e.\ $\theta_D$ of glassy TiZrNb alloy, and the knowledge of $\theta_D$ is the prerequisite for the determination of the value of $\lep$, thus the type of superconductivity of an alloy \cite{KUVEZDIC2020119865,McMillanPR1968}.
However, if we assume that an approximately linear variation of $\lep$ with $N_0(E_F)$ observed in \Ni\ alloys (see Figure 6 in \cite{KUVEZDIC2020119865}) extends to higher values of $N_0(E_F)$ we can estimate that $\lep \approx 0.66$ corresponds to $N(E_F)$ (Figure \ref{fig:gamma and nef}) of a glassy TiZrNb alloy.
This value of $\lep$ corresponds to an intermediate coupled BCS superconductor.

\subsection{Upper critical magnetic field $\Hcz$ and coherence length $\xi(0)$}
\begin{figure}
\centering
\includegraphics[width=0.6\linewidth]{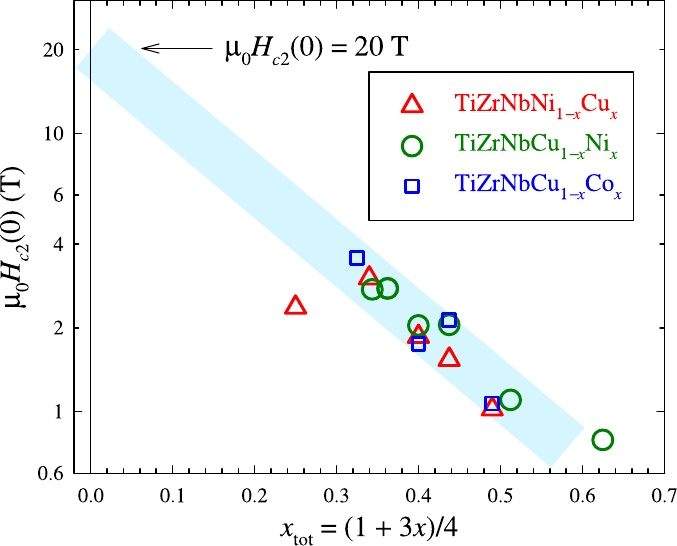}
\caption{ 
The upper critical field $\Hcz$ of \Cu, \Ni\ and \Co\ glassy alloys (red triangles, green circles and blue squares, respectively) for various $x$. 
Data are adapted from Refs.\ \cite{KUVEZDIC2020119865,ma16041711}.
Thick light blue line denotes linear extrapolation to $\xtot \rightarrow 0$, i.e.\ to glassy TiZrNb alloy.
Arrow denote extrapolated value of upper critical field $\Hcz = 20\,$T.\label{fig:Hc2}
}
\end{figure}

The knowledge of $\Hc$ is important both for understanding and practical applications of superconductors \cite{met10081078,KUVEZDIC2020119865,Tafra_2008,tinkham2004introduction}.
In particular, the value of $\Hcz = \Hc(T=0)$ may provide an insight into the mechanism of destruction of superconductivity and can also indicate whether the studied system can be described as a conventional (BCS) superconductor or not.
The values of $\Hcz$ of HEA and CCA superconductors are often very high \cite{met10081078,KUVEZDIC2020119865, Kosut2024LTP,SunPRM2019,Li_2024}, exceeding the Pauli paramagnetic limit $\mu_0 H_P = 1.83 T_c$ \cite{ClogstonPRL1962}.
Since the values of $\Hcz$ of all our \all\ glassy alloys were close to $H_P$ \cite{KUVEZDIC2020119865,ma16041711} we show the variation of their values with $\xtot$ using the same type of plot as that for $T_c$ in Figure \ref{fig:Tc}.
As seen in Figure \ref{fig:Hc2}, the values of $\Hcz$ of all our alloys (except for the slightly crystalline alloy with \pue{0}{Cu} \cite{Ristic2019JAP}) follow the same, approximately linear variation, which for $\xtot = 0$ extrapolates to that of the glassy TiZrNb alloy, $\mu_0\Hcz = (20 \pm 5)\,$T.
This value is somewhat higher than $\mu_0 H_P = (15 \pm 2)\,$T and is a little lower than the highest value of $\mu_0\Hcz$ reported so far for HEAs \cite{wu_polymorphism_2020}.
As discussed in Ref.\ \cite{ma16041711}, the definition of $\Hc$ using the midpoint of the resistive transition would, due to broadening of the resistive transition in an applied magnetic field, result in a somewhat lower values of $\Hcz$ in Figure \ref{fig:Hc2}, thus a lower value of $\Hcz$ of glassy TiZrNb alloy.
However, rather sharp resistive transitions observed in all our samples \cite{KUVEZDIC2020119865,ma16041711} would result in the extrapolated value of $\Hcz$ which is still within the extrapolated errors in Figure \ref{fig:Hc2}, thus still somewhat higher than $\mu_0 H_P$ corresponding to the $T_c$ of the glassy TiZrNb alloy.
Considerably lower values of $\Hcz$, $12\,$T \cite{Kosut2024LTP} or $13\,$T \cite{GABANI2023414414}, have been reported for crystalline TiZrNb alloy.
These values were estimated using the WHH approximation in the dirty limit \cite{WerthamerPR1966}:
\begin{equation}
    \label{eq:WHH}
    \mu_0\Hcz = -0.69 T_c 
    \left(\mu_0 \frac{d\Hc}{dT}\right)_{T_c}.
\end{equation}

Despite some ambiguity associated with the application of WHH approximation to $\Hc$ results covering a narrow magnetic field, $\mu_0 \Hc \le 5\,$T and temperature range $T \ge 0.7 T_c$, and the absence of data confirming the dirty limit, we expect that $\Hcz$ value of the crystalline TiZrNb alloy will be lower than that of the corresponding amorphous alloy.
Indeed, a strong topological disorder of the amorphous phase makes the electronic mean free path very short (thus high electrical resistivity \cite{ma16041711}).
This results in a short superconducting coherence length and the associated enhancement of $\Hc$ \cite{Kittel2005,tinkham2004introduction}.
Indeed, large ratio of $\Hcz/T_c$ in V-Nb-Mo-Al-Ga HEAs over an extended composition range was attributed to disorder-induced enhancement of $\Hc$ \cite{wu_polymorphism_2020}.
Unfortunately, a rather large scatter of the values of electrical resistivity $\rho$ in our \all\ alloys \cite{KUVEZDIC2020119865,ma16041711} prevent sufficiently accurate extrapolation of these results to that of TiZrNb glassy alloy (thus, an estimate of the amount of disorder in this alloy).
However, from the average slopes of the decrease of $\rho$ with decreasing TL contents we can make a rough estimate of $\rho$ in amorphous phase TiZrNb, $\rho_a \gtrsim 100\,\mu\Omega$cm, which is about two or more times larger than the values of $\rho$ in similar crystalline refractory HEAs and CCAs \cite{SunPRM2019}.

\begin{figure}
\centering
\includegraphics[width=0.6\linewidth]{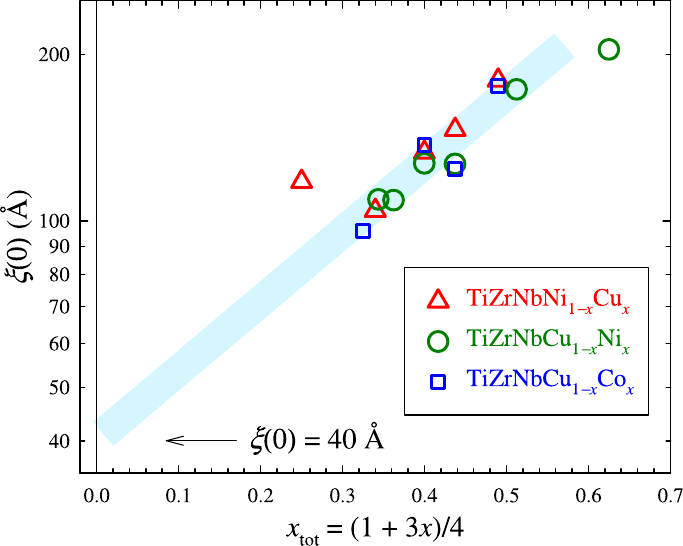}
\caption{ 
The coherence length $\xi(0)$ of \Cu, \Ni\ and \Co\ glassy alloys (red triangles, green circles and blue squares, respectively) for various $x$.
Thick light blue line denotes linear extrapolation to $\xtot \rightarrow 0$, i.e.\ to glassy TiZrNb alloy.
Arrow denote extrapolated value of coherence length $\xi(0) = 40\,$\AA.\label{fig:ksi}
}
\end{figure}

Using the Ginzburg-Landau expression \cite{Kittel2005,tinkham2004introduction} we have calculated the coherence length $\xi(0)$ of all \all\ glassy alloys using the corresponding values of $\mu_0\Hcz$ from Figure \ref{fig:Hc2}:
\begin{equation}
    \label{eq:ksi}
    \xi(0) = 
    \left(
        \Phi_0/2\pi \mu_0 \Hcz
    \right)^{0.5}.
\end{equation}
Here, $\Phi_0$ is the magnetic flux quantum $\Phi_0 = h/2e$, where $h$ is the Planck constant and $e$ is the charge of the electron.
(We note that due to simple proportionality between $T_c$ and $\mu_0\Hcz$ in superconductor with Pauli paramagnetic limit \cite{ClogstonPRL1962} one can also use $T_c$ in order to estimate $\xi(0)$.)
As seen in Figure \ref{fig:ksi}, $\xi(0)$ values of our alloys decrease approximately exponentially with decreasing $\xtot$ (the exception is a large $\xi(0)$ of the partially crystalline alloy with zero Cu content), and extrapolate to $\xi(0) = (40 \pm 3)\,$\AA\ for TiZrNb glassy alloy.
This value of $\xi(0)$ agrees very well with that obtained by the use of $\mu_0\Hcz = (20 \pm 5)\,$T in Eq.\ (\ref{eq:ksi}), $\xi(0) = (40.6 \pm 3.5)\,$\AA\ which seems to indicate relatively good accuracy of our extrapolation procedures. 
The values of $\xi(0)$ of the best conventional superconductors Nb$_3$Sn and NbTi of around $30\,$\AA and $50\,$\AA, respectively \cite{BANNO2023100047}, are close to that estimated for the glassy TiZrNb superconductor.
The coherence length is the fundamental length scale in superconducting state which affects all thermodynamic and electromagnetic properties of a superconductor \cite{Kittel2005,tinkham2004introduction}, thus affects both the understanding and practical applications of superconductivity.
As one example we note that the magnitude of $\xi(0)$ determines the size of the optimal flux pinning centers, thus also the magnitude of $J_c$ in an applied filed (e.g.\ \cite{Novosel_2012}).
Recent finding of the chemical inhomogeneities associated with the affinity between Ti and Zr atoms in the crystalline TiZrNb alloy \cite{XUN2023221} with the nanometer size may indicate rather strong intrinsic pinning of magnetic vortices at these inhomogeneities, thus rather high values of the in-field $J_c$.

\subsection{VEC and $T_c$ relationship in TE-TL alloys}
\begin{figure}
\centering
\includegraphics[width=0.6\linewidth]{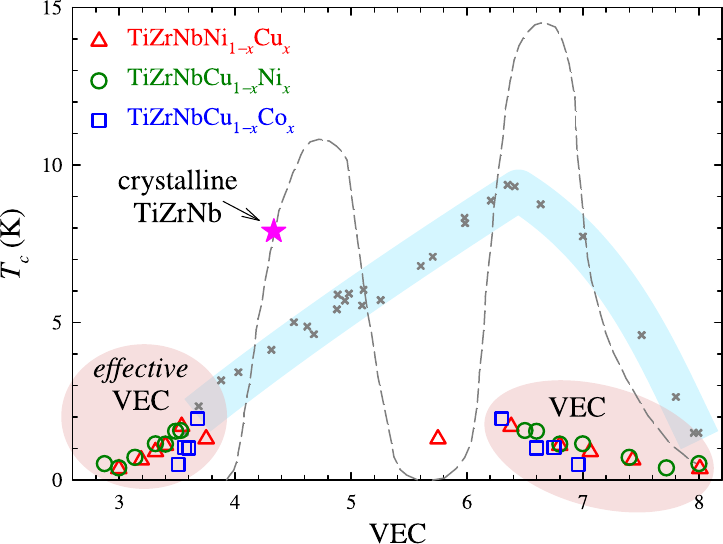}
\caption{
The superconductivity transition temperature $T_c$ dependence on VEC for \Cu, \Ni\ and \Co\ glassy alloys, and other TE-TL alloys (both HEAs and the conventional one). 
Gray crosses corresponds to data for amorphous alloys, gray dashed line corresponds to data for classic crystalline alloys (data are adapted from Refs.\ \cite{CollverPRL1973,MatthiasPR1955}).
Light blue band is guide to the eye for amorphous alloys data.
Pink star symbol denotes (single) value of $T_c$ for crystalline TiZrNb \cite{Kosut2024LTP}.
Red triangles, green circles and blue squares correspond to \Cu, \Ni and \Co, respectively (data are adapted from Refs.\ \cite{KUVEZDIC2020119865,ma16041711}); on the right side of the plot is their dependence on VEC; on left side of the plot is their dependence on \textit{effective} VEC (see text).  
\label{fig:VEC}
}
\end{figure}

At the end we will briefly address the much debated problem of the relationship between VEC and $T_c$ in TE-TL alloys, both HEAs and the conventional one \cite{met10081078,KUVEZDIC2020119865,ma16041711,BABIC1981139,RISTIC2010S194,Tafra_2008,SunPRM2019,XU2024115986,AltounianPRB1983,KarkutPRB1983,StolzeCoM2018}.
Already the first systematic study of the superconductivity and magnetism of glassy Zr-TL alloys \cite{BABIC1981139,babic_conference_1982} showed that the variation of $T_c$ with VEC does not follow those of either amorphous thin films of neighboring 4d transition metals (light blue band and corresponding gray crosses in Figure \ref{fig:VEC}, \cite{CollverPRL1973}) or their crystalline counterparts (gray dashed line in Figure \ref{fig:VEC}, \cite{MatthiasPR1955}).
In Zr-Ni,Cu alloys \cite{BABIC1981139,RISTIC2010S194,AltounianPRB1983,KarkutPRB1983,babic_conference_1982} $T_c$, $\chiexp$ and $N(E_F)$ decrease continuously with increasing Ni,Cu content (thus VEC) instead of increasing towards $\text{VEC} \simeq 6.5$ as in amorphous thin films of neighboring 4d metals (Figure \ref{fig:VEC} and \cite{CollverPRL1973}).
This `anomalous' variation of $T_c$ with VEC in amorphous TE-TL alloys is simply explained by their split-band structure of the DOS within the valence band, observed in photoemision spectra (PES) of both crystalline and amorphous TE-TL alloys (see e.g.\ \cite{AMAMOU19801029,OELHAFEN19801017,ZEHRINGER1988317}) and confirmed by subsequent calculation of their ES \cite{PhysRevB.27.2049,Jank_1991,HAFNER1992307}.
PES and theory have shown that within the superconducting range of compositions, $N(E_F)$ is dominated by the d-states of TEs and the contribution of those of TLs to $N(E_F)$ is quite small \cite{AMAMOU19801029,OELHAFEN19801017,ZEHRINGER1988317,ma16041486}.
Accordingly, $T_c$ has to decrease with increasing TL content and with conventionally defined VEC ($\text{VEC} = \sum_ic_i\text{VEC}_i$, where $c_i$ and $\text{VEC}_i$ are the molar fraction and VEC of the $i$-th component, respectively) which overestimates the contribution of d-states of TLs to $N(E_F)$.
Indeed, by using an \textit{effective} VEC for $\text{TL} = \text{Ni}$ and Cu, equal to their valency (thus substantially lower than the total number of their valence electrons) it was possible to reconcile the variation of $T_c$ with \textit{effective} VEC in glassy Zr-Cu(Ni) alloys \cite{ma16041486} with that of amorphous alloys of neighboring 4d metals \cite{CollverPRL1973}.
The discovery of HEA and CCA superconductors lead to the (re)discovery of the discrepancy between the variation of $T_c$ with VEC predicted in either \cite{CollverPRL1973} or \cite{MatthiasPR1955} and that observed in TE-TL composed HEAs and CCAs \cite{met10081078,SunPRM2019}.
Initially, despite the observed decrease of $N(E_F)$ with increasing TL content (see e.g.\ \cite{StolzeCoM2018}), this discrepancy was not understood and was believed to be specific to chemical complexity of HEAs.
The PES studies of TE-TL based HEAs and CCAs, both glassy and crystalline, have shown qualitatively the same structure of DOS within the valence band as in their binary and ternary counterparts (see Ref.\ \cite{ma16041486} and references therein).
Thus their variations of $T_c$ with VEC can also be reconciled with these observed in alloys of neighboring 4d metals \cite{CollverPRL1973,MatthiasPR1955} providing that one uses the \textit{effective} VEC for TLs (which is physically better justified).
This is demonstrated in Figure \ref{fig:VEC} which shows that the erroneous variation of $T_c$s of our alloys with conventional values of VEC (Figure \ref{fig:VEC}, right side) is corrected with the use of \textit{effective} $\text{VEC} = 2$ for Cu and Ni, respectively (Figure \ref{fig:VEC}, left side).
Further, we show that $T_c$ of the crystalline TiZrNb alloy \cite{Kosut2024LTP} obeys the Matthias rule \cite{MatthiasPR1955} (pink star symbol in Figure \ref{fig:VEC}).

\section{Conclusion}

We performed a novel analysis of \replaced{existing}{the} results of a comprehensive study of the changes in physical properties of three quinary amorphous alloy systems: \Cu, \Ni\ and \Co\ on transition from high entropy alloy to conventional alloy regime.
\added{In our opinion,} our analysis provided reliable estimates of the physical properties of amorphous TiZrNb alloy which has not been prepared in the amorphous phase yet.
\replaced{Most of the properties of glassy TiZrNb would be similar to those of crystalline TiZrNb refractory alloy, except for certain superconducting properties, in praticular higher values of the upper critical field, which is probably consequence of the strong topological disorder of the amorphous phase.
This opens up the amorphous TiZrNb as a possible basis for the contemporary refractory alloys and might be also a promising biomedical alloy.}{Most of the properties of glassy TiZrNb are similar to those of crystalline TiZrNb refractory alloy which is the basis for the contemporary refractory alloys and is also a promising biomedical alloy.} 
\replaced{Moreover}{However}, the glassy form of TiZrNb alloy could outperform the crystalline one in biomedical applications and in superconducting properties in an applied magnetic field.
An explanation of the discrepancy between the variations of $T_c$ with the average number of valency electrons \replaced[id=EB]{accross neighboring}{in} alloys of 4d transition metals and some high entropy alloys has been provided.

The presented method of analyzing systematic results is rather general and can provide reliable estimates of the properties of any alloy which has not been prepared as of yet.
Moreover, our analysis enables a better description of the properties of the single-element metallic glasses than that based on actual samples.

\begin{acknowledgments}
This work was partially supported by the Croatian Science Foundation Projects No. IP-2022-10-6321 HOI2DEM. We acknowledge the support of project CeNIKS co-financed by the Croatian Government and the European Union through the European Regional Development Fund, Competitiveness and Cohesion Operational Programme (Grant no. KK.01.1.1.02.0013). We also acknowledge the support of the project Cryogenic Centre at the Institute of Physics -- KaCIF co-financed by the Croatian Government and the European Union through the European Regional Development Fund-Competitiveness and Cohesion Operational Programme (Grant No. KK.01.1.1.02.0012) and the support of the project Ground states in competition – strong correlations, frustration and disorder -- FrustKor financed by the Croatian Government and the European Union -- NextGenerationEU through the National plan for recovery and resilience 2021. -- 2026.  Financial support from the University of Zagreb through funds for multipurpose institutional financing of scientific research (project 104-F24-00011) is greatly acknowledged.
\end{acknowledgments}

\bibliography{biblio}

\end{document}